\documentstyle[prl,aps,preprint]{revtex}
\input epsfig.sty

\def\dk{\frac{d^4k}{(2 \pi)^4}}
\def\dthreek{\frac{d^3k}{(2 \pi)^3}}

\def\D{\Delta}

\def\ep{\epsilon}

\def\dia{{\rm \ diagram}}

\begin{document}
\setcounter{page}{0}
\begin{normalsize}
\begin{flushright}
      MIT-CTP-2490\\
      UR-1447\\
      MRI-Phy-95-26\\
      hep-ph/9603325\\
\end{flushright}
\end{normalsize}
\def\footnoterule{\kern-3pt \hrule width\hsize \kern3pt}
\tighten
\begin{LARGE}
\begin{center}
{CUTTING RULES AT FINITE TEMPERATURE}
\end{center}
\end{LARGE}
\vspace{.5cm}
\begin{center}
Paulo F. Bedaque\footnote{Email address: {\tt bedaque@ctpa04.mit.edu}}

\vspace{2mm}       

Center for Theoretical Physics \\
Laboratory for Nuclear Science \\
and Department of Physics \\
Massachusetts Institute of Technology \\
Cambridge, Massachusetts 02139 \\

\vspace{.5cm}

Ashok Das

\vspace{2mm}

Department of Physics and Astronomy\\
University of Rochester\\
Rochester, N.Y. 14627\\

\vspace{.5cm}

Satchidananda Naik\footnote{Email address: {\tt naik@mri.ernet.in}}

\vspace{2mm}
Mehta Research Institute of Mathematics\\
and Mathematical Physics\\
10, Kasturba Gandhi Marg\\
Allahabad 211002, India\\

\vspace{2mm}
( February 1996)\\
\end{center}

\thispagestyle{empty}

\begin{center}
{\bf Abstract}
\end{center}

We discuss the cutting rules in the real time approach to finite temperature
field theory and show the existence of  cancellations among classes of cut 
graphs which allows a physical interpretation of the imaginary part of the
relevant amplitude in terms of underlying microscopic processes. Furthermore, 
with these cancellations, any calculation of the imaginary part of an amplitude
becomes much easier and completely parallel to the zero temperature case.

\vspace*{\fill}

\section{Introduction}
At zero temperature, the evaluation of the imaginary part of a scattering
amplitude is immensely simplified by the use of the cutting (Cutkosky) rules
\cite{ref:cutkosky}
which further make the unitarity of the S-matrix manifest. The fact that the
imaginary part of a $n$-loop amplitude can be related to the on-shell
amplitudes with lower order loops is best seen, in the modern language, 
by the use of the so called largest time equation of 
't Hooft and Veltman~\cite{ref:diagrammar}. Generalizations  of these
rules to the finite temperature case are essential and useful since a large  
number of transport coefficients are given by the imaginary part of
some equilibrium finite temperature correlators. The cutting rules at finite
temperature will not only simplify the evaluation of these coefficients but 
will also help in understanding the structure of field theories at finite
temperature as well as identifying the microscopic processes underlying them.
In this paper, we provide such a generalization in the real time approach to
finite temperature field theories.

Finite temperature (and density) extensions of the cutting rules
have been discussed, in the past, both in the imaginary 
time formalism \cite{ref:jeon} as well as in the real time formalism 
\cite{ref:semkob}. Explicit imaginary time  calculations of the self-energy 
by Weldon,
in various theories, 
led him to identify the 
physical meaning of the imaginary part in
terms of underlying microscopic processes.
On the other
hand, the attempt by Kobes and Semenoff to generalize the cutting rules in the
real time formalism ran into difficulties beyond the one loop. (Their
discussion was completely within the framework of Thermofield Dynamics.) The
main reason for this discrepancy appeared to be the fact that large classes of
diagrams that vanish at zero temperature (the graphs that cannot be written as
cut diagrams) do not vanish at finite temperature due to the distinct form of
the finite temperature propagators,
and this extra class of diagrams do not admit the physical interpretation
sugested by Weldon.
Such new graphs arise only at two and
higher loops. 
More recently, Jeon has proposed Cutkosky-like rules for the computation 
of imaginary parts in the imaginary time approach. It also includes
graphs that can not be interpreted as  cut diagrams.
 This raises two possible interesting scenarios. Namely, either the
physical interpretation of the imaginary parts 
or the cutting rules proposed
need to be analyzed more carefully.

In this paper we discuss the generalization of the cutting rules for the finite
temperature case in the real time approach. Our discussion is within the
context of the Closed Time Path formalism where the propagators satisfy various
identities which makes the analysis of complicated diagrams a lot easier. We
show that while graph by graph there arise new contributions (distinct from the
zero temperature case), large classes of graphs cancel among themselves leading
to the fact that there are no ``extra'' diagrams to be considered different from
the zero temperature case. The cancellations arise because of the properties of
the propagators as well as the KMS conditions present in the theory. This
shows, in particular, that the interpretation sugested by Weldon, in fact, holds and
that the imaginary part of an amplitude at any loop can be written as a sum
over cut diagrams much like the zero temperature case. This has the added
advantage that the number of diagrams needed to evaluate the imaginary part
reduces considerably and it makes clear the physical meaning of this in terms
of underlying microscopic processes. The paper is organized as follows. In
section II, we introduce the notion of circling and discuss the generalization
of the cutting rules. In section III, we prove the cancellation among classes
of diagrams leading to the correct description of the imaginary part in terms
of cut diagrams. In section IV, we discuss this explicitly at one and two loops
with the example of a $\phi^{3}$ theory alongwith the physical interpretation
and present a short conclusion in section V.

\section{Cutting rules}

In the real time formalism, the number of field degrees of freedom double at
finite temperature leading to a $2\times 2$ matrix structure for the Greens
functions and the propagators of the theory. In the Closed Time Path formalism,
for example, we can write the Greens functions of, say, a scalar theory as
($a,b=+,-$)

\begin{equation}
\D_{ab} = \left(\begin{array}{cc}
                   \D_{++} & \D_{+-}\\
                    \D_{-+} & \D_{--}
                 \end{array}\right)
\end{equation}
where in momentum space, the Greens functions have the form

\begin{mathletters}%
\label{eq:prop}
\begin{eqnarray}
\D_{++}(k)&=&  
        {1\over k^2-m^2+i\epsilon} - 2\pi i n(k_0)\delta(k^2-m^2)\ ,
          \label{eq2:a}\\
              \D_{--}(k)&=& 
        -{1\over k^2-m^2-i\epsilon} - 2\pi i n(k_0)\delta(k^2-m^2)\ ,
          \label{eq2:b}\\
               \D_{+-}(k)&=& 
        -2\pi i (n(k_0) + \theta(-k_0))\delta(k^2-m^2)\ ,
           \label{eq2:c}\\
               \D_{-+}(k)&=& 
        - 2\pi i (n(k_0) + \theta(k_0))\delta(k^2-m^2)\ ,
           \label{eq2:d}  
\end{eqnarray}
\end{mathletters}%
with $n(k_0)$ representing the bosonic distribution function at $\beta=1/kT$

\[ n(k_0) = {1\over e^{\beta |k_0|}-1}. \] 
It is interesting to note from the explicit forms of the Greens functions above
that they satisfy the identity

\begin{equation}
\label{eq:sumzero}
\D_{++} + \D_{--} = \D_{+-} + \D_{-+}.
\end{equation}
We can now arrive at the finite temperature generalization of the cutting rules
following closely the argument presented in ~\cite{ref:diagrammar}.
The K\"{a}llen-Lehman spectral representation for the Green's function
$\D_{ab}$, at finite temperature, has the form ~\cite{ref:abrikosov}

\begin{equation}
\D_{ab} (x)=\int^\infty_0\ ds\ \int \dk \left[ 
       {\rho_{ab}(s,\vec k)\over k^2 - s + i \ep}+
       {\tilde\rho_{ab}(s,\vec k)\over k^2 - s - i \ep}\right] e^{-ik\cdot x} ,
\end{equation}
where $\rho_{ab}(s,k),\tilde\rho_{ab}(s,k)$ are the spectral functions.
We can perform  the integral over $k_0$ and it is easy now
to see that $\D_{ab}(x)$ 
can be written as

\begin{equation}
\D_{ab}(x)=\theta(x^0) \D^+_{ab}(x) +\theta(-x^0) \D^-_{ab}(x) ,
\end{equation}
where the functions $\D^\pm_{ab}(x)$ are defined by

\begin{equation}
\label{eq:Dasfunctionofrho}
\D^\pm_{ab}(x)= \int^\infty_0\ ds\ \int \dk \ 2\pi i \delta(k^2-m^2)
         \left[ - \theta(\pm k_0) \rho(s,\vec k) +  \theta(\mp k_0) \tilde\rho(s,\vec k)
         \right] e^{-ik\cdot x}.
\end{equation}

We note here that the spectral functions can be read out from the structure of
the propagators in momentum space to be
\begin{mathletters}%
\begin{eqnarray}
\label{eq:7}
\rho_{++}(s,k)&=&\delta(s-m^2) (1+n(k_0)), \quad 
\tilde\rho_{++}(s,k)=-\delta(s-m^2) n(k_0)\\
\rho_{--}(s,k)&=&\delta(s-m^2) n(k_0), \quad 
\tilde\rho_{--}(s,k)=-\delta(s-m^2) (1+n(k_0))\\ 
\rho_{+-}(s,k)&=&-\tilde\rho_{+-}(s,k)=\delta(s-m^2) (\theta(-k_0)+n(k_0)), \\ 
\rho_{-+}(s,k)&=&-\tilde\rho_{-+}(s,k)=\delta(s-m^2) (\theta(k_0)+n(k_0)). 
\end{eqnarray}
\end{mathletters}%
The fact that $\rho(s,k)=-\tilde\rho(s,k)$ for the ($+-$) and ($-+$) functions
reflects the fact that those functions, being solutions of the homogeneous
equations of motion, are regular at $x^0$.
This brings out the distinctive feature at finite temperature, namely, that,
unlike the case at zero temperature, the functions $\D^\pm_{ab} (x)$ contain 
both positive and negative frequencies. This is what leads to additional
contributions at finite temperature to the imaginary part of an amplitude graph
by graph.

Next, we define the notion of circling much the same way as at zero
temperature. A propagator with one of the ends circled is defined by replacing  
$\D_{ab}(x-y)$ with $\D^+_{ab}(x-y)$ if the vertex $x$ is circled and with
$\D^-_{ab}(x-y)$ if the vertex $y$ is circled. A propagator with both the ends
circled, namely, if both $x$ and $y$ are circled, is defined by replacing 
$\D_{ab}$ by 
$\tilde\D_{ab}(x-y)=\theta(x^0)\D^-(x-y) + \theta(-x^0)\D^+(x-y)$  . 
In addition we also assign
an extra factor of $-1$ to each circled vertex. These rules are summarized
in figure 1.

\begin{center} 
\epsfig{file=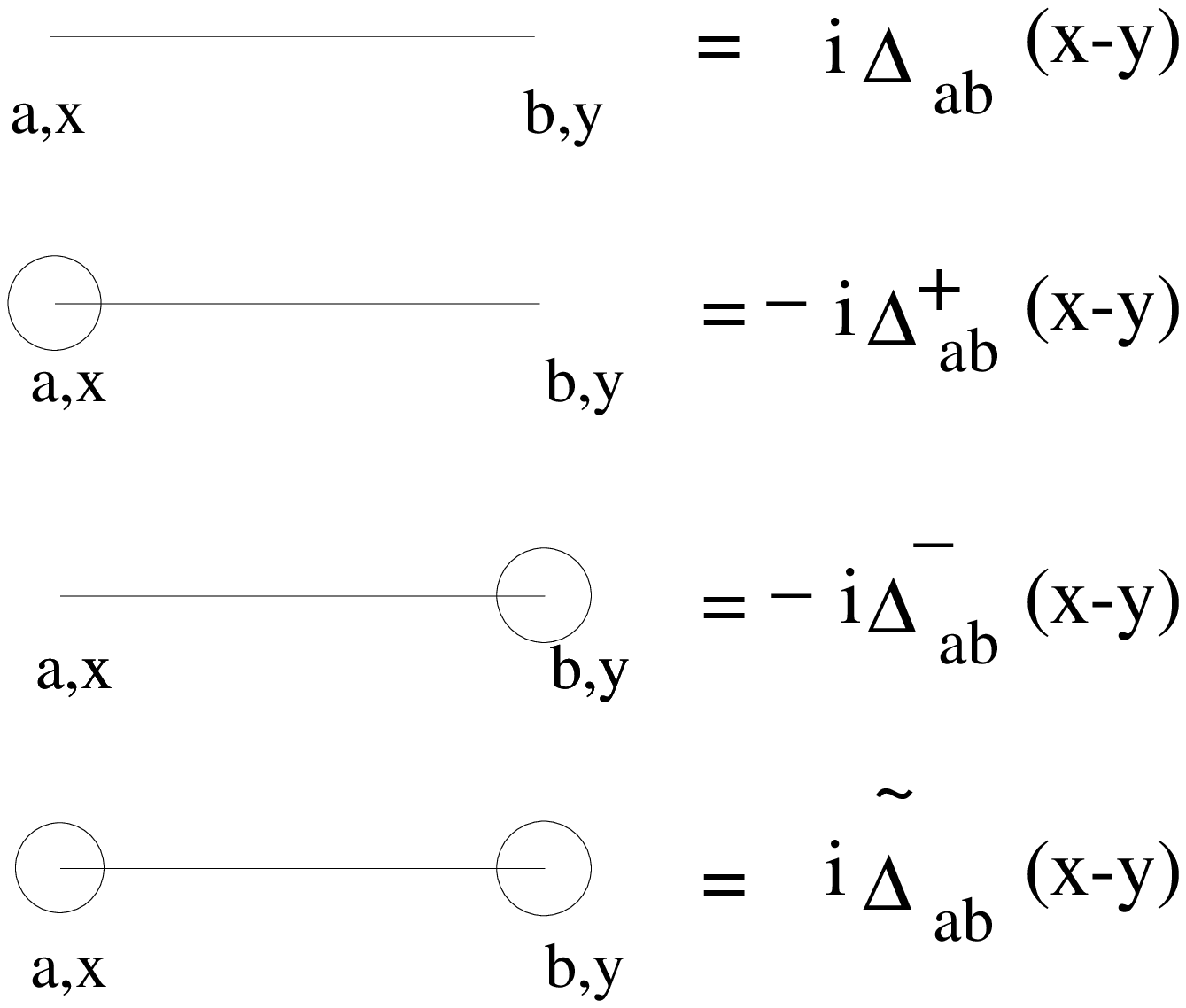,height=8 truecm}

{\large Figure 1}
\end{center}
\bigskip

More explicitly, the four components of the propagator of the theory,
$\D_{++},\D_{--},\D_{+-}$ and $\D_{-+}$, have the following
circled representations (see equations \ref{eq:Dasfunctionofrho}
and \ref{eq:7}).

\begin{center} 
\epsfig{file=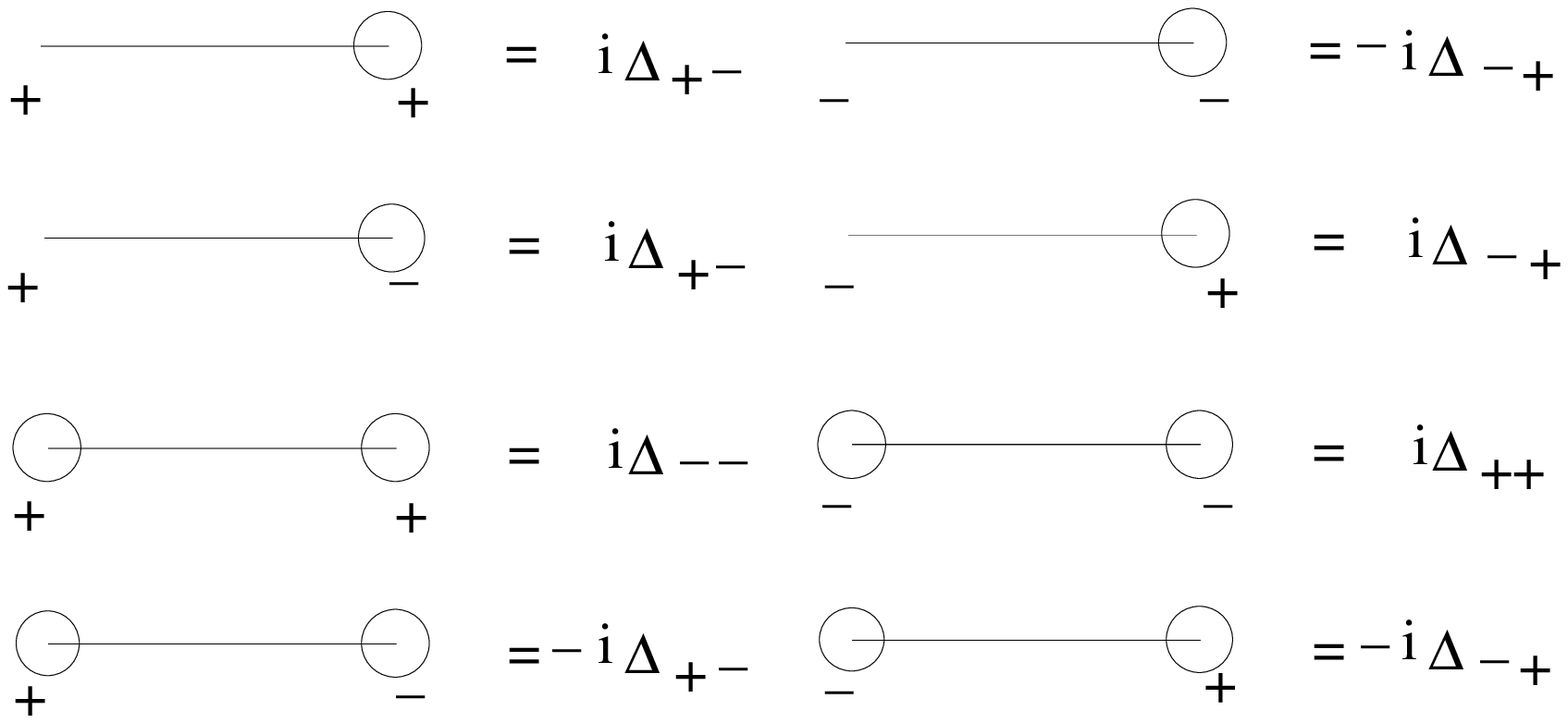, height=8 truecm}

{\large Figure 2}
\end{center}
\bigskip

In addition to the identity in eq. (\ref{eq:sumzero}), the propagators at finite temperature
satisfy other relations as well. Thus, for example, the KMS condition would say
that
\begin{equation}
\D_{ab}(t,\vec{x}) = \D_{ab}(t-i\beta,\vec{x})
\end{equation}
which, in momentum space, leads to
\begin{eqnarray}
\label{eq:kms}
\D_{+-}(k) & = & e^{-\beta k_0} \D_{-+}(k)\nonumber\\
\D_{-+}(k) & = & e^{\beta k_0} \D_{+-}(k).
\end{eqnarray}
Furthermore, from  the definitions, it is easy to show that
\begin{equation}
\label{eq:cc}
\big( i \D_{ab}\big)^*(p) = i \tilde\D_{ab}(p)
\end{equation}
We note that these are only some of the relations that will be useful in our
analysis. The propagators satisfy various other relations which can be obtained
from their definitions.

From these discussions, we arrive at the largest time equation much like at
zero temperature. The largest time equation says that
a graph with vertices  $x_1,...,x_n$ plus the same graph with the vertex
with the largest time circled is zero. (It is easy to see from the definitions
that there exists a corresponding smallest time equation as well, but we will
not use it in our analysis.)

\begin{equation}
\label{eq:largetime}
G(x_1,...,x_k,...,x_n)+G(x_1,...,{\bf x_k},...,x_n)=0, \ \ 
x^0_k > x^0_1,...x^0_n,
\end{equation}
where a bold face vertex stands for a circled vertex.
Equation (\ref{eq:largetime}) is a direct consequence of the definitions
of figure 1.
Notice that (\ref{eq:largetime}) is true regardless of whether other
vertices besides $x_k$ are circled.  
From (\ref{eq:largetime}) it now follows that the sum of any graph over
all possible circlings must vanish. Namely,

\begin{equation}
G(x_1,...,x_k,...,x_n)+G({\bf x_1},...,{\bf x_k},...,{\bf x_n})+
\sum_{{\rm circlings}} G(x_1,...,x_k,...,x_n) = 0,
\end{equation}
\noindent
where the sum, in the last term, is over all possible circlings excluding the
diagram where all  vertices are circled.
To show this, we observe that the sum above can be decomposed  into a sum over 
pairs of diagrams with a given set of circlings and with the largest time
vertex circled or uncircled.  By (\ref{eq:largetime}) these pairs vanish and,
therefore, the sum adds up pairwise to zero. Using (\ref{eq:cc}) , we now 
see that we can write the above relation also as 

\begin{equation}
{\rm Im}\quad iG(p_1,...,p_m)=\frac 12(iG(p_1,...,p_m)+(iG(p_1,...,p_m))^*)
                        =-\frac 12\sum_{circlings}iG(p_1,...,p_m).
\label{eq:im}
\end{equation}

At zero temperature the functions $\D^+(k)$ ($\D^-(k)$) contain only
positive (negative) energies and, using energy conservation at each vertex,
it is possible to show that a circled vertex cannot be surrounded by uncircled
vertices only (and {\it vice versa}). In other words, such graphs vanish by
energy conservation. (Incidentally, this is how the unphysical degrees of
freedom of the finite temperature field theory drop out of physical amplitudes
at zero temperature.) As a result, the only nontrivial graphs are those where
the circled vertices form a connected
region in a diagram that includes one external vertex (that is, a vertex 
connected 
to an external line). This leads to the notion of cutting which basically says
that in such a case, it is possible to separate the graph into a circled and an
uncircled region by drawing a line through the graph (see  figure 3).

\begin{center} 
\epsfig{file=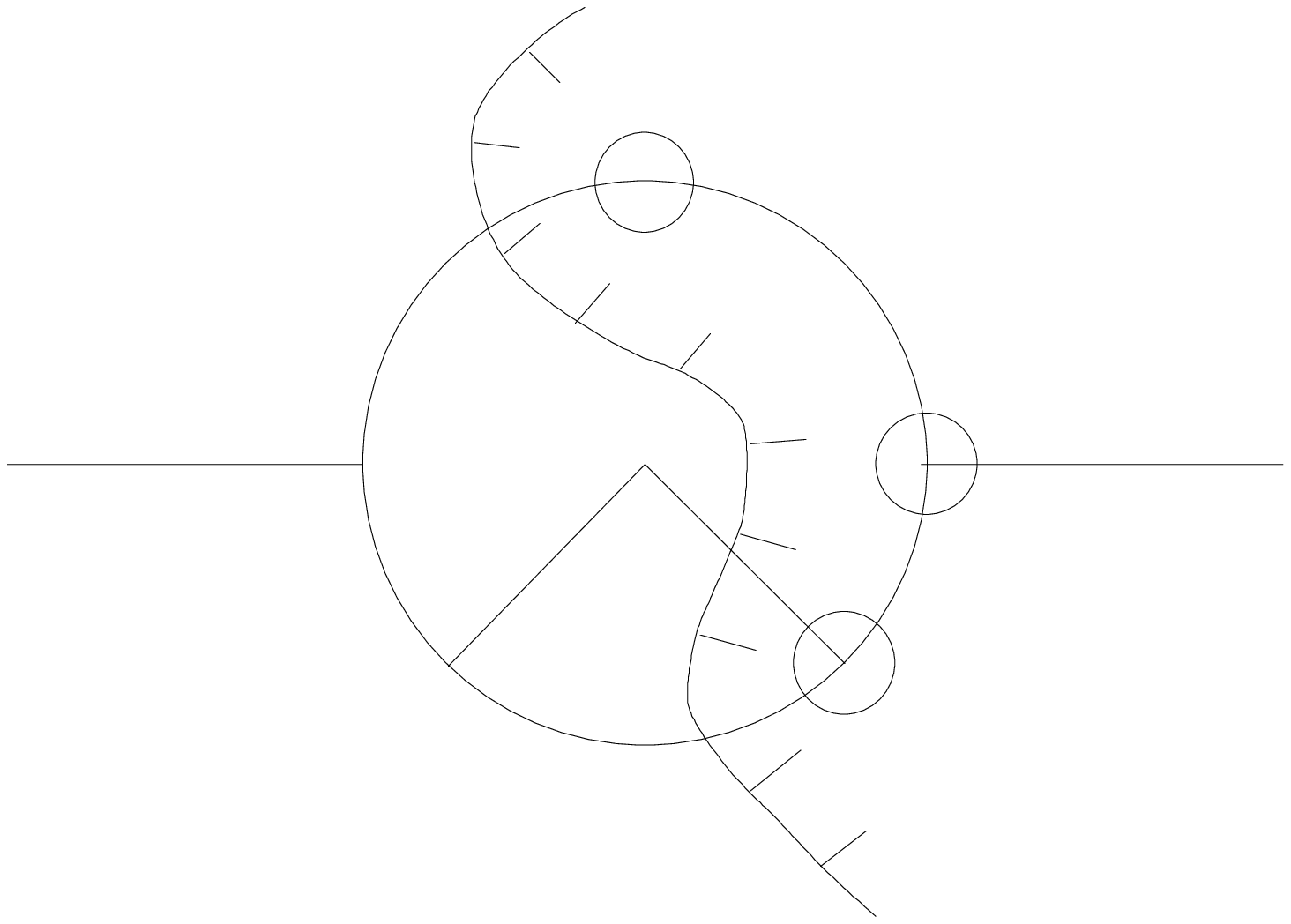,height=6.5 truecm}

{\large Figure 3}
\end{center}
\bigskip
\noindent The last term 
in (\ref{eq:im}), in this notation, simply becomes a sum over all cut diagrams
and, in this case, equation  (\ref{eq:im}) leads to the usual zero temperature
Cutkosky rule.

On the other hand, as we mentioned earlier, at finite temperature
the functions $\D^\pm(k)$ contain both  positive and negative frequency 
components and as a result, graphs with isolated circled vertices or isolated 
uncircled vertices do not vanish anymore. These contribute and  should be 
included in the right hand side of (\ref{eq:im}). These are the additional
graphs that we alluded to earlier and it is not clear any more whether in the
presence of these new graphs, the notion of cutting would go through. In the
next section, we will argue that even though individually such graphs
contribute, there exist classes of such graphs which cancel to leave us with no
unwanted diagrams.

\section{ Cancellations among  graphs}

The computation of the imaginary part of any Green's function
involves a double sum at finite temperature, namely, we have to sum
over all possible circlings as in zero temperature, but in addition we must
also sum over the two kinds of intermediate vertices $+$ and $-$ because of the
doubled degrees of freedom. Thus, at finite temperature, for any Greens
function, we have 
\begin{equation}
\label{eq:double}
{\rm Im}\ iG_{a...b}=-\frac 12\sum_{{\rm internal\  \  vertices}=\pm}\ 
\sum_{{\rm circlings}}\ \dia,
\end{equation}
where $a...b=\pm$ are the ``thermal indices'' of the external legs.
As we have mentioned before, unlike the zero tempertaure case,
circlings that cannot be represented by
a cut diagram as in figure 2 also appear in (\ref{eq:double}).
Examples of such graphs are in figure 6(i,j,k,l,m,n).
All such graphs, however, are combinations of the same basic propagators 
$\D_{++},\D_{--},\D_{+-}$ and $\D_{-+}$
regardless of the combination of $+$ and $-$ vertices, circled or
uncircled. We will show now that, after summing
over the internal thermal index $+$ and $-$,
there are cancellations among such graphs which makes  unnecessary
to consider unwanted kinds of circlings.  The circlings that we are  left  
with finally, are of the kind of those in figure 2, which can be 
represented by a cut 
diagram.
This not only simplifies the evaluation of the imaginary parts by reducing
the number of graphs to be considered but also
allows a physical interpretation in terms of deacy rates and emission/absorption
of particles to/from the medium.

To proceed, let us note that any self-energy graph with any combination of 
circled and uncircled vertices can be drawn in one of the three forms shown 
in figure 4(a,b,c).

\begin{center} 
\epsfig{file=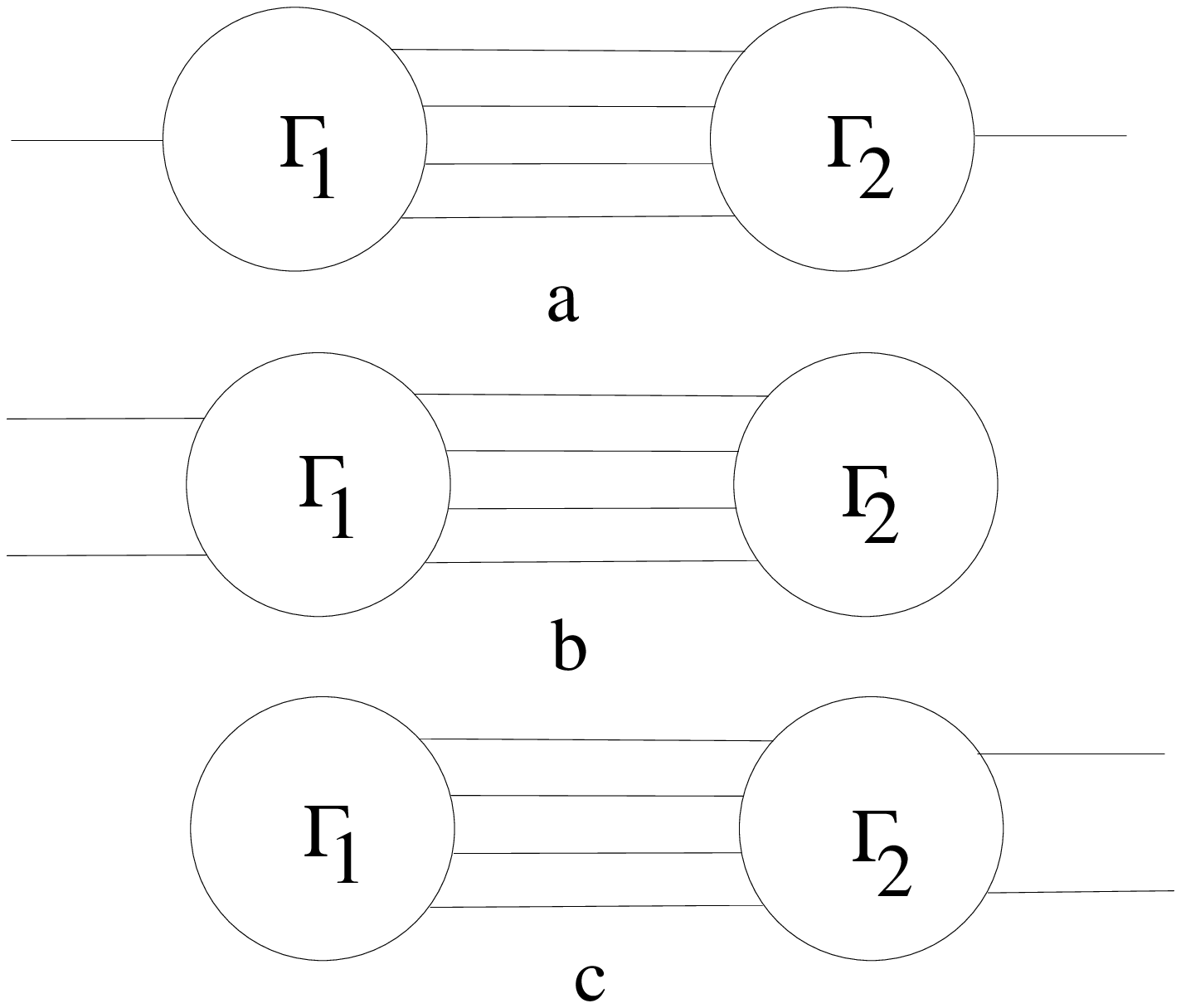,height=10 truecm}

{\large Figure 4}
\end{center}
\bigskip\noindent
Here, $\Gamma_1$ is assumed to contain
all the circled vertices of the graph and only the circled vertices while 
$\Gamma_2$ contains all the  uncircled vertices of the graph and only the 
uncircled ones. The vertices in $\Gamma_1$ (or $\Gamma_2$) need not all be
connected to each other.
The kind of diagrams that have a non vanishing contribution to the right hand side of
(\ref{eq:im}) at {\it zero temperature}  are the ones that can be drawn
as in figure 4(a), with  all the vertices in $\Gamma_1$ connected with
each other, as well with all vertices of $\Gamma_2$ connected with each other.
We say that graphs of this form can be drawn as a cut graph since
a cut between $\Gamma_1$ and $\Gamma_2$ separates all circled vertices
from the uncircled ones, as well as the incoming from the outgoing lines, leaving
no island of circled or uncircled vertices isolated from external lines.
On the other hand, graphs such as the ones in figure 5(b,c) cannot
be represented by a cut diagram since one cannot draw a line separating a circled
from uncircled vertices which will also separate the incoming and the  outgoing 
external lines. Graphs like
figure 4(a) with $\Gamma_1$ (or $\Gamma_2$) disconnected
can not be represented by a cut diagram either, since a line separating 
$\Gamma_1$ from $\Gamma_2$ will also split $\Gamma_1$ ($\Gamma_2$) in two,
leaving an cluster of circled (uncircled) vertices isolated from
any external line.
We will now show that the diagrams that can not be represented by a cut diagram,
in the sense explained above,
vanish after a summation of the indices $+$ and $-$ of the internal
vertices is performed. This amounts 
to saying that the kinds of circlings that we need to consider, at finite
temperature, are precisely the ones appearing at zero temperature.
To show this,  we essentially need two main results. 

Our first result is that any graph containing a connected cluster of 
{\it circled} vertices 
not attached to
any external line must vanish.
This  result   disposes of the graphs
of the form of figure 4(c) as well as the ones of the form of figure 5(a)
if $\Gamma_1$ does not form a   connected set of vertices. 
We can see that by  taking this connected cluster of circled vertices
isolated from external lines  to be $\Gamma_1$ in graphs like    figure 4(c),
or the component of $\Gamma_1$ disconnected from the external lines in graphs
like figure 4(a). 

This result is most easily proven in position space, using
a variation of the argument presented in \cite{ref:diagrammar}.
Let us consider the cluster of circled vertices isolated from the 
external lines that
we assume to exist and 
let us focus on the  circled vertex {\it inside this cluster} with 
the {\it smallest}
time of all the vertices in the cluster.
Since it is assumed that the external 
vertices are not circled, this must be
an internal vertex and, as such, we should sum over its thermal index
$a=\pm$. In general, this vertex will be connected to both $+$ and $-$ 
vertices, 
circled and uncircled, as well with itself (forming tadpoles). Let us assume
that this smallest time vertex (among the circled vertices) is connected to 
$n$ uncircled  type $+$  vertices at coordinates $y_1,...,y_n$,
$m$ uncircled  type $-$  vertices located at $z_1,...,z_m$, $p$ circled 
type $+$  vertices located at $r_1,...,r_p$, $q$ circled type $-$  vertices
at $s_1,...,s_q$ and to itself $l$ times (see figure 5). 

Summing over the thermal index of the smallest time circled vertex, and
taking into account that

\begin{mathletters}
\begin{eqnarray}
\Delta_{++}(x)&=&\theta(x^0) \Delta_{-+}(x)+\theta(-x^0) \Delta_{+-}(x)\\
\Delta_{--}(x)&=&\theta(x^0) \Delta_{+-}(x)+\theta(-x^0) \Delta_{-+}(x),
\end{eqnarray}
\end{mathletters}
\noindent
we have

\begin{center} 
\epsfig{file=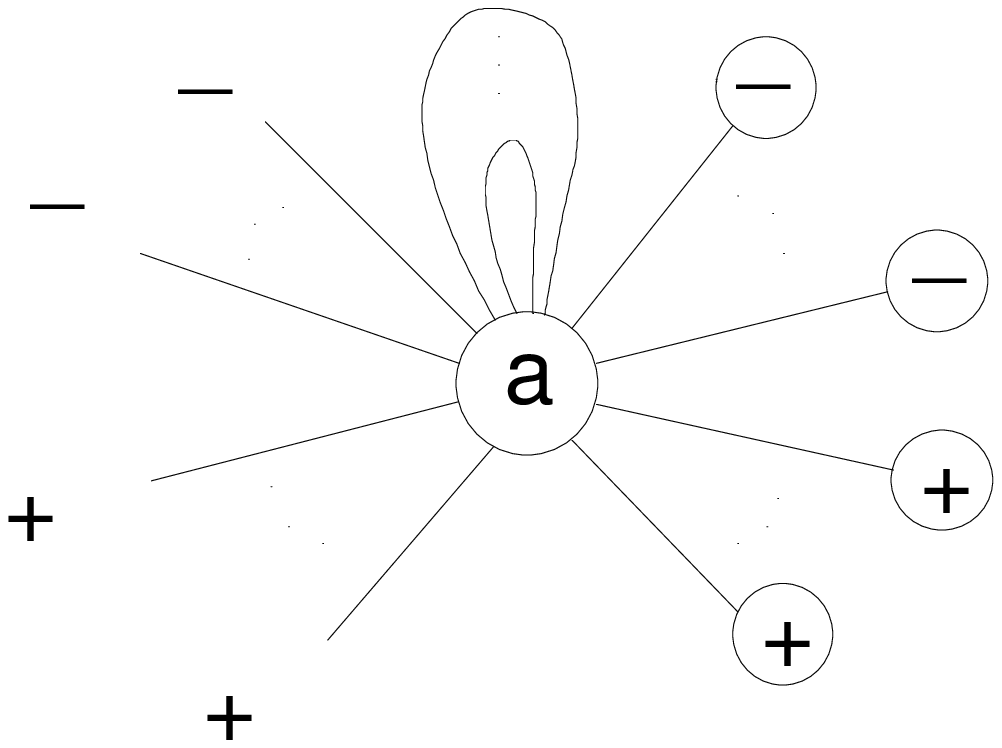,height=7 truecm}

{\large Figure 5}
\end{center}
\bigskip

\begin{eqnarray}
\sum_{a=\pm} \dia =& R\ \D_{--}^l(0)& \D_{-+}(x-y_1)...\D_{-+}(x-y_n)
                                      \D_{+-}(x-z_1)...\D_{+-}(x-z_m)\nonumber\\
                                   && \D_{--}(x-r_1)...\D_{--}(x-r_p)
                                      \D_{+-}(x-s_1)...\D_{+-}(x-s_q)\nonumber\\
                  &-R\ \D_{++}^l(0)&  \D_{-+}(x-y_1)...\D_{-+}(x-y_n)
                                      \D_{+-}(x-z_1)...\D_{+-}(x-z_m)\nonumber\\
                                   && \D_{-+}(x-r_1)...\D_{-+}(x-r_p)
                                      \D_{++}(x-s_1)...\D_{++}(x-s_q)\nonumber\\
                  =&R\ \D_{++}^l(0)&  \D_{-+}(x-y_1)...\D_{-+}(x-y_n)
                                      \D_{+-}(x-z_1)...\D_{+-}(x-z_m)\nonumber\\
                  &&\big[            \D_{-+}(x-r_1)...\D_{-+}(x-r_p)
                                      \D_{+-}(x-s_1)...\D_{+-}(x-s_q)\nonumber\\
                  &&                  \D_{-+}(x-r_1)...\D_{-+}(x-r_p)
                                      \D_{+-}(x-s_1)...\D_{+-}(x-s_q)
                    \big]\nonumber\\
                  =&0,\hfill&
\end{eqnarray}
where $R$ stands for the rest of the diagram which is not connected to the
circled vertex with the smallest time (and contains a multiplicative factor of
$i^{m+n+p+q+l}$ coming from the definition of the propagators. In addition, 
we have used the fact that 
\begin{equation}
\D_{++}(0)=\D_{--}(0),
\end{equation}
to arrive at the second equality. We should add here that even though the
coordinates of the intermediate vertices are being integrated over, this result
still holds because there will always be one circled intermediate vertex with
the smallest time among the group and the proof would go through.
 
This proves the first of the two results mentioned above, namely, the sum 
over the thermal indices of the 
internal vertices vanishes if there is a connected cluster of circled 
vertices which does not include the external vertices. 
As we have pointed out earlier, this includes  graphs of the form of 
figure 4(c) as well as figure 4(a) for disconnected $\Gamma_1$.

It is worth stressing here that this result does not apply to graphs containing
only one cluster of circled vertices which include at least one external
vertex. This is because, in such a case, for some values of the coordinates of
the intermediate vertices, the external circled vertex may, in fact, be the one
with the smallest time. On the other hand, being an external vertex, its index
is fixed for a particular amplitude and should not be summed and as a result
the theorem fails. A further cancellation, however, occurs in the computation
of the imaginary part of the retarded self-energy given by $\Sigma_R =
\Sigma_{++} - \Sigma_{+-}$. It is clear that evaluation of this would involve,
in addition to summing over the thermal indices of the internal vertices, a sum
over the thermal index of one of the external vertices (examples are the ones in figure
6(e,f,g,h)). As a result, these graphs should vanish when the external index
being summed is a circled vertex by the above theorem.

\begin{center} 
\epsfig{file=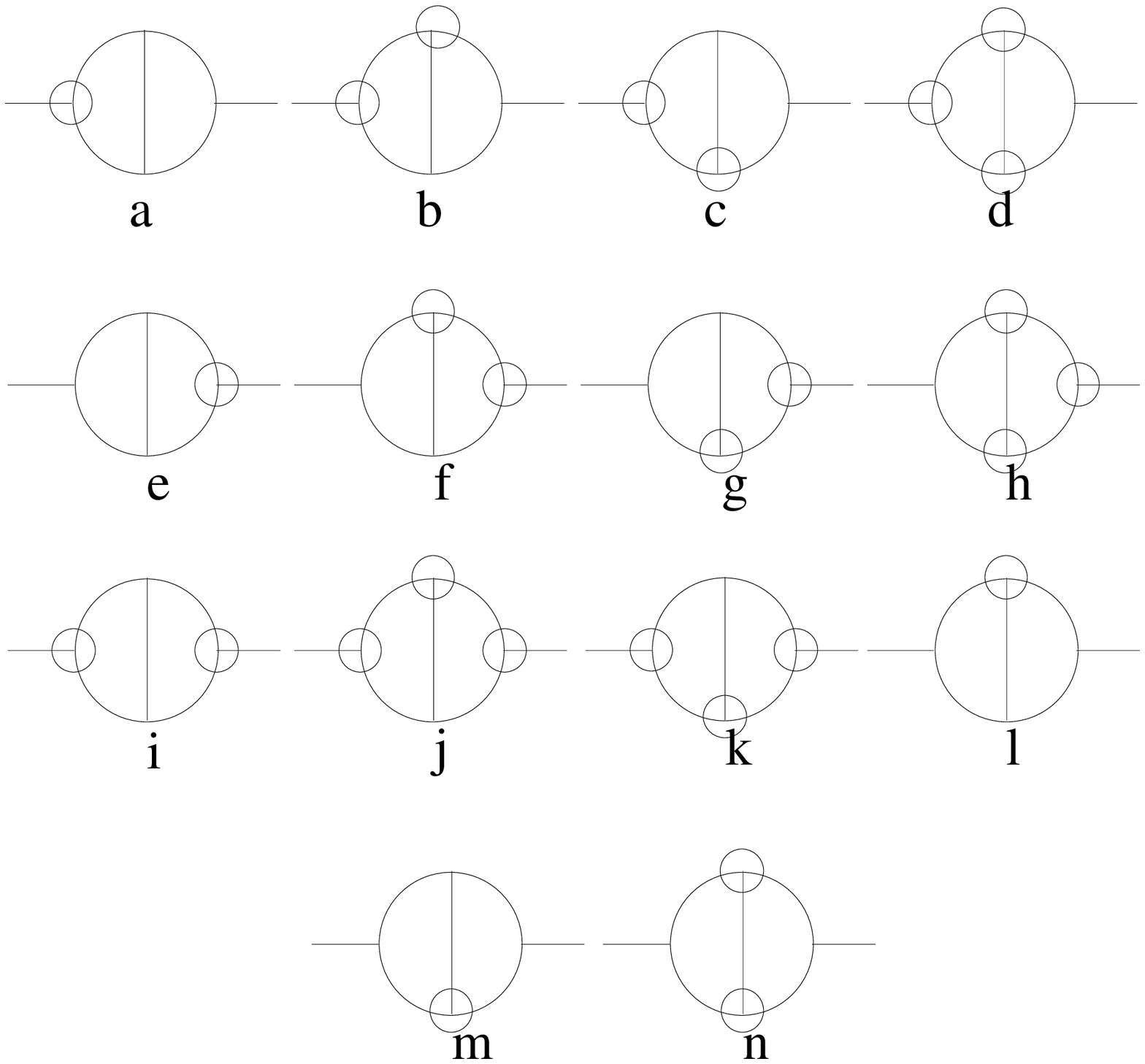,height=12 truecm}

{\large Figure 6}
\end{center}
\bigskip

The second result which we need is,
in a sense, complementary to  the first one. We will  show that graphs containing
a connected cluster of {\it uncircled} vertices  not connected to any 
external line vanishes. 
This result implies that graphs like those of figure 4(b), as well that those graphs
like the figure 4(a) with $\Gamma_2$ composed of two or more disconnected pieces, 
must vanish. To see this we just take this connected cluster of uncircled vertices
not attached to any external line to be $\Gamma_2$, in graphs like
figure 4(b),  or
a component of $\Gamma_2$ not attached to the external line, in graphs like figure
4(a).

We can show our second result in the following way. Let  
$k_1, ..., k_n$ represent the momenta carried by the lines leaving 
$\Gamma_1$ and going  into  the cluster of uncircled vertices
not attached to any external line $\Lambda$ ( $\Lambda$ is $\Gamma_2$ for graphs of the 
form of figure 4(b),
but represents only the component of $\Gamma_2$ not attached to any external vertices for
graphs like the figure 4(a) with $\Gamma_2$ disconnected).
The important thing to observe here is that the propagators connecting 
$\Gamma_1$ and $\Lambda$ do not depend on the thermal indices of the vertices in
$\Gamma_1$, but only on the vertices on $\Lambda$. This is because these are
necessarily propagators with only one end circled and as is clear from 
figure 2 in this case, the propagators have opposite thermal indices labelled
completely by the uncircled vertex. Let us consider next a graph of the type
of figure 4(b) (or of the type 4(a) with $\Gamma_2$ disconnected)
 for an arbitrary fixed set of thermal indices for the 
vertices that are not in $\Lambda$ while we sum over the thermal indices of 
the $\Lambda$.  For any such configuration, these graphs  
factorize as

\begin{equation}
\Sigma(p)= \Gamma(p, k_1,...,k_n) \sum_{\alpha_1,...,\alpha_n=\pm}
           \Lambda_{ \alpha_1,...,\alpha_n}( k_1,...,k_n)
           \Delta_{\alpha_1\ -\alpha_1}(k_1)...\Delta_{\alpha_n\ -\alpha_n}(k_n).
\end{equation}
where   $\Lambda_{ \alpha_1,...,\alpha_n}( k_1,...,k_n)$  and $\Gamma (p,k_1,...,k_n)$
represent
the contributions coming from connecting the vertices in  and outside of $\Lambda$ 
 respectively (as well as factors of $i$ coming from the propagators
and vertices). The sum over each index $\alpha_i$ of the
vertices in $\Lambda_{ \alpha_1,...,\alpha_n}( k_1,...,k_n)$ can be performed as 
follows

\begin{eqnarray}
\sum_{\alpha_i=\pm}\Lambda_{~ \ ... \alpha_i ...~}( k_1,...,k_n)
                   \ ...\Delta_{\alpha_i\ -\alpha_i}(k_i)\ ...\ &=&
                   \Lambda_{\ \ ... +  ...}( k_1,...,k_n)
                   \ ...\Delta_{-+}(k_i)...\nonumber  \\
                 &\ +&\Lambda_{~ \ ... -  ...}( k_1,...,k_n)
                   \ ...\Delta_{+-}(k_i)...\ \\
                  &=&\ ...\Delta_{+-}(k_i)...\ [\Lambda_{\ \ ... +  ...}( k_1,...,k_n)
                                                e^{\beta k_i^0}\nonumber\\
                  &\phantom{=} &\phantom{
                               \ ...\Delta_{+-}(k_i)...\
                                        }
                                +\Lambda_{~  ... -  ...}( k_1,...,k_n)\nonumber
                                          ], 
\end{eqnarray}
where we used the KMS condition of eq.(\ref{eq:kms})

\begin{equation}
\Delta_{-+}(k)=e^{\beta k_0}\Delta_{+-}(k),
\end{equation}
to rearrange some of the propagators.
The signs associated with the $+$, $-$ vertices as well as the circlings are
included in the definition of the $\Gamma$'s . After the sum over all the 
indices $\alpha_1,...,\alpha_n$ we are left with

\begin{equation}
\label{eq:sigmasum}
\Sigma(p)= \Gamma(p, k_1,...,k_n)
           \Delta_{+-}(k_1)...\Delta_{+-}(k_n)
           \sum_{\alpha_1,...,\alpha_n=\pm}
           \Lambda_{ \alpha_1,...,\alpha_n}( k_1,...,k_n)
           e^{\beta \sum_{j=1}^n {\alpha_j+1 \over 2} k_j^0}.
\end{equation}

Let us note here that  ${\alpha_j+1 \over 2}=1$ or $0$ depending on whether 
$\alpha_j=1$ or $-1$. The terms in the sum above combine pairwise to produce

\begin{eqnarray}
\label{eq:inter}
&\Lambda_{\alpha_1,...,\alpha_n}( k_1,...,k_n)
e^{\beta \sum_{j=1}^n {\alpha_j+1 \over 2} k_j^0}+
\Lambda_{-\alpha_1,...,-\alpha_n}( k_1,...,k_n)
e^{\beta \sum_{j=1}^n {-\alpha_j+1 \over 2} k_j^0}\nonumber\\
&\phantom{\Lambda_{\alpha_1,...,\alpha_n}( k_1,...,k_n)
e^{\beta \sum_{j=1}^n {\alpha_j+1 \over 2} k_j^0} }
 = \Lambda_{\alpha_1,...,\alpha_n}^*( k_1,...,k_n)+
\Lambda_{-\alpha_1,...,-\alpha_n}^*( k_1,...,k_n).
\end{eqnarray}

This is easily seen as follows. Let us note that 
$\Lambda (k_1,...,k_n){\alpha_1,...,\alpha_n}$
represents a graph
containing only uncircled vertices of type $+$ and $-$. Therefore, let us
consider a generic (amputated) diagram
with $n$ ($+$) external lines, $m$ ($-$) external lines, and draw it by grouping the
positive (negative) internal vertices into ``blobs'' $\Lambda_+^n$ 
($\Lambda_-^m$), as in figure 7. 

\begin{center}
\epsfig{file=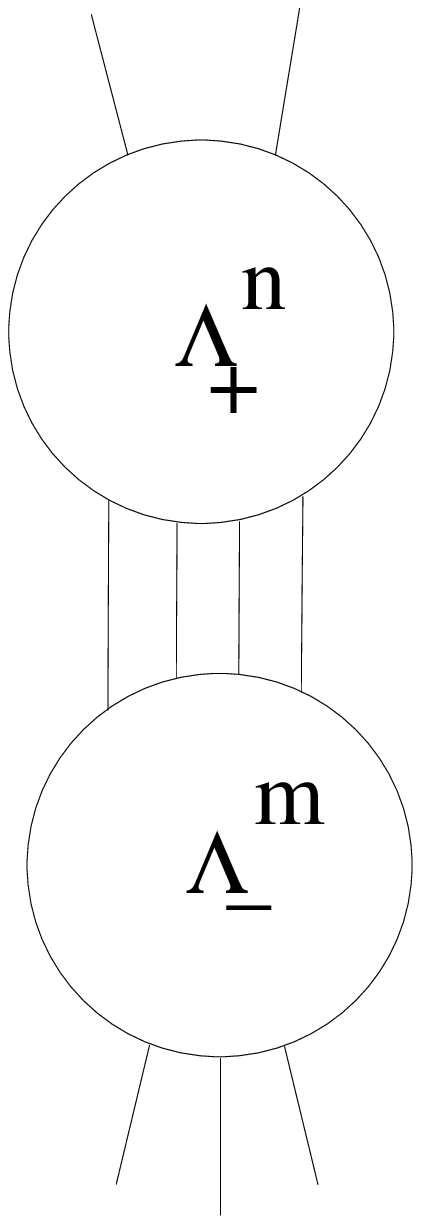,height=8 truecm}

{\large Figure 7}
\end{center}
\bigskip\noindent
Taking the complex conjugate of this diagram
(in momentum space) and using the fact that $ (i\Delta_{++})^*=i\Delta_{--})$ and
$(i\Delta_{\pm\mp})^*=i\Delta_{\pm\mp})$ as well as  the relation \ref{eq:kms}, 
we find that (see figure 8) for any number of $+$ and $-$ indices

\begin{equation}
\label{eq:kmsgen}
\Lambda_{+...+\ -...-}^*=\Lambda_{-...-\ +...+} e^{-\beta p_0},
\end{equation}
where $p$ is the total momentum flowing through $\Gamma_{+...+\ -...-}$, 
namely, it is the total momentum entering into the diagram of figure 7 through
the $+$ vertices. Using this result and
combining $\Lambda_{\ \ \alpha_1,...,\alpha_n}$ and 
$\Lambda_{\ \ -\alpha_1,...,-\alpha_n}$ 
we now arrive at (\ref{eq:inter}).

\begin{center}
\epsfig{file=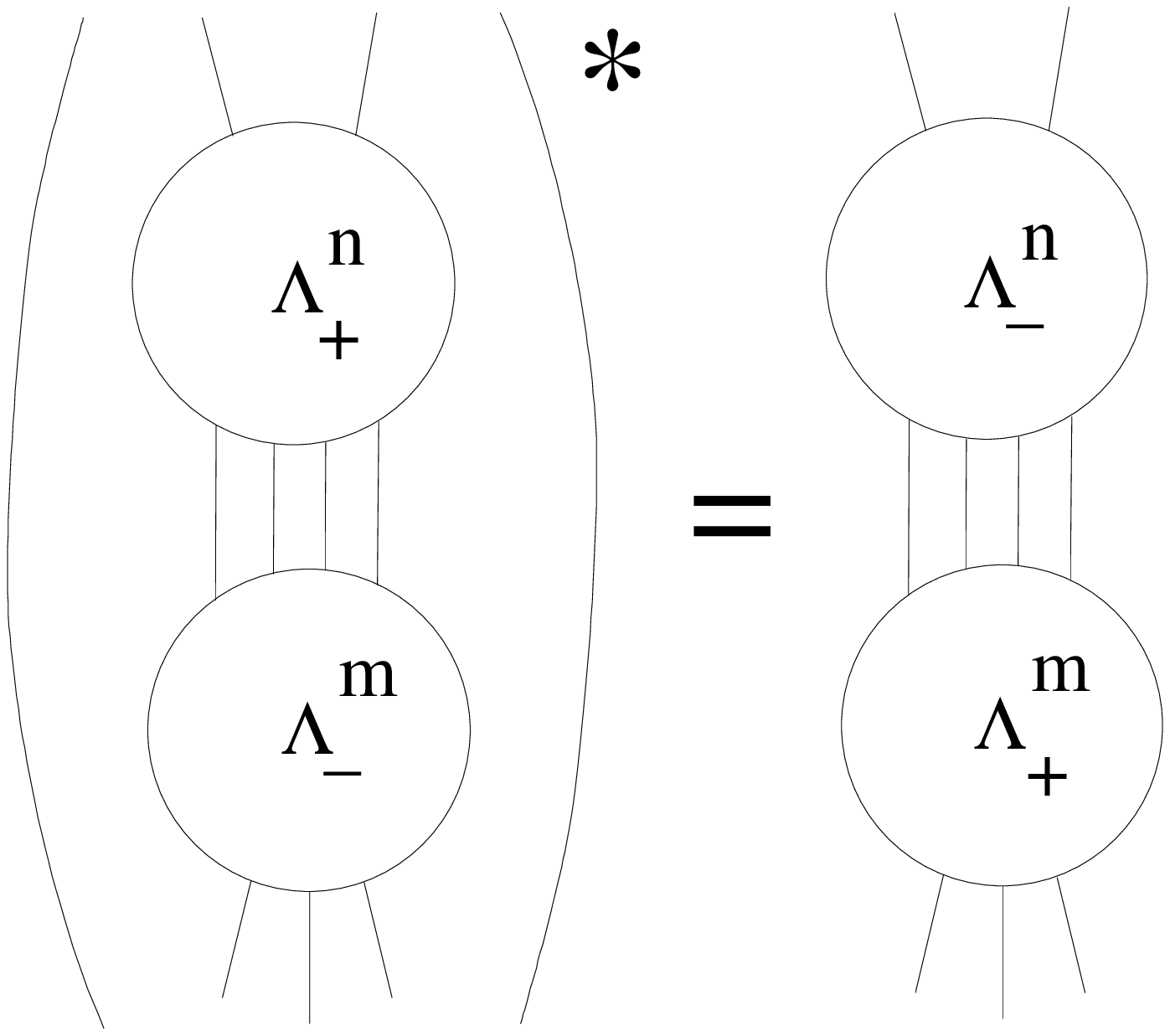,height=8 truecm}

{\large Figure 8}
\end{center}
\bigskip\noindent
The equation (\ref{eq:kmsgen})  can be regarded as a generalization of the 
KMS condition for higher point functions. Returning to equation 
(\ref{eq:sigmasum}) we now have

\begin{equation}
\Sigma(p)= \Gamma(p, k_1,...,k_n)\Delta_{+-}(k_1)...\Delta_{+-}(k_n)
           \sum_{\alpha_1,...,\alpha_n=\pm}
           \Lambda_{\ \ \alpha_1,...,\alpha_n}^*( k_1,...,k_n). 
\end{equation}

We can now show easily that the sum above vanishes using the largest/smallest
time argument.
Let us consider a graph contributing to $\Lambda_{\alpha_1,...,\alpha_n}$
in position space and choose the vertex with the largest time labelling
its coordinate and thermal index respectively as $x$ and $a$. 
This vertex is connected generically
to $n$ type $+$  vertices located at $y_1,...,y_n$ and
$m$ type $-$  vertices located at $z_1,...,z_n$ (there are no circled
vertices in $Lambda$). The sum over $a=\pm$ is easily seen to give

\begin{eqnarray}
{\rm diagram}& = &  R [ \Delta_{++}(x-y_1)...\Delta_{++}(x-y_n)
                  \Delta_{+-}(x-z_1)...\Delta_{+-}(x-z_n)\nonumber\\ 
              &\ &  -\Delta_{-+}(x-y_1)...\Delta_{++}(x-y_n)
                  \Delta_{--}(x-z_1)...\Delta_{--}(x-z_n)]\nonumber\\
      &  = &  R [\Delta_{-+}(x-y_1)...\Delta_{-+}(x-y_n)
                  \Delta_{+-}(x-z_1)...\Delta_{+-}(x-z_n)\nonumber\\ 
              &\ & -\Delta_{-+}(x-y_1)...\Delta_{-+}(x-y_n)
                  \Delta_{+-}(x-z_1)...\Delta_{+-}(x-z_n)]\nonumber\\
      & = &  0,
\end{eqnarray}
where $R$ is the remaining of the diagram, independent of the vertex on $x$.

This completes the proof of the result alluded to earlier, namely,  graphs 
containing a cluster of uncircled vertices not attached to any external line vanishes.

With these two results, then, we see that the cutting rules for the imaginary
part of any diagram has the same form as at zero temperature. The unwanted
"extra" graphs present at finite temperature cancel among themselves when we
sum over the thermal indices of the internal vertices. Thus, for example, it is
clear now that of all the graphs in figure 6, the only ones that will give a
nontrivial contribution to the evaluation of the imaginary part of the 
the retarde self-energy are 4(a,b,c,d) \footnote{The imaginary part
of the causal self energy is just a multiple of the imaginary part of
the retarded self energy
\cite{ref:abrikosov}}.

\section{Physical Interpretation}

Let us illustrate the cutting rule for the imaginary part of a diagram,
discussed in the last section, with a very simple example. We will consider
the imaginary part of the one loop  retarded self-energy for a 
scalar theory with a $\phi^3$ interaction. As we have mentioned earlier, the
retarded and the causal self-energies are related by a multiplicative factor
and yet we choose to examine the retarded function only because this is the one
that has a direct connection with the calculations in the imaginary time
formalism. Furthermore, it is the retarded function at finite temperature where
the cutting of a diagram leads in a simple way to the unitarity relation. The
imaginary part of the retarded self-energy for this theory is 
given, at one loop, by the graphs in figure 9 (the graphs with the
 vertex on the right circled  cancel) and an easy calculation gives
 
\begin{center} 
\epsfig{file=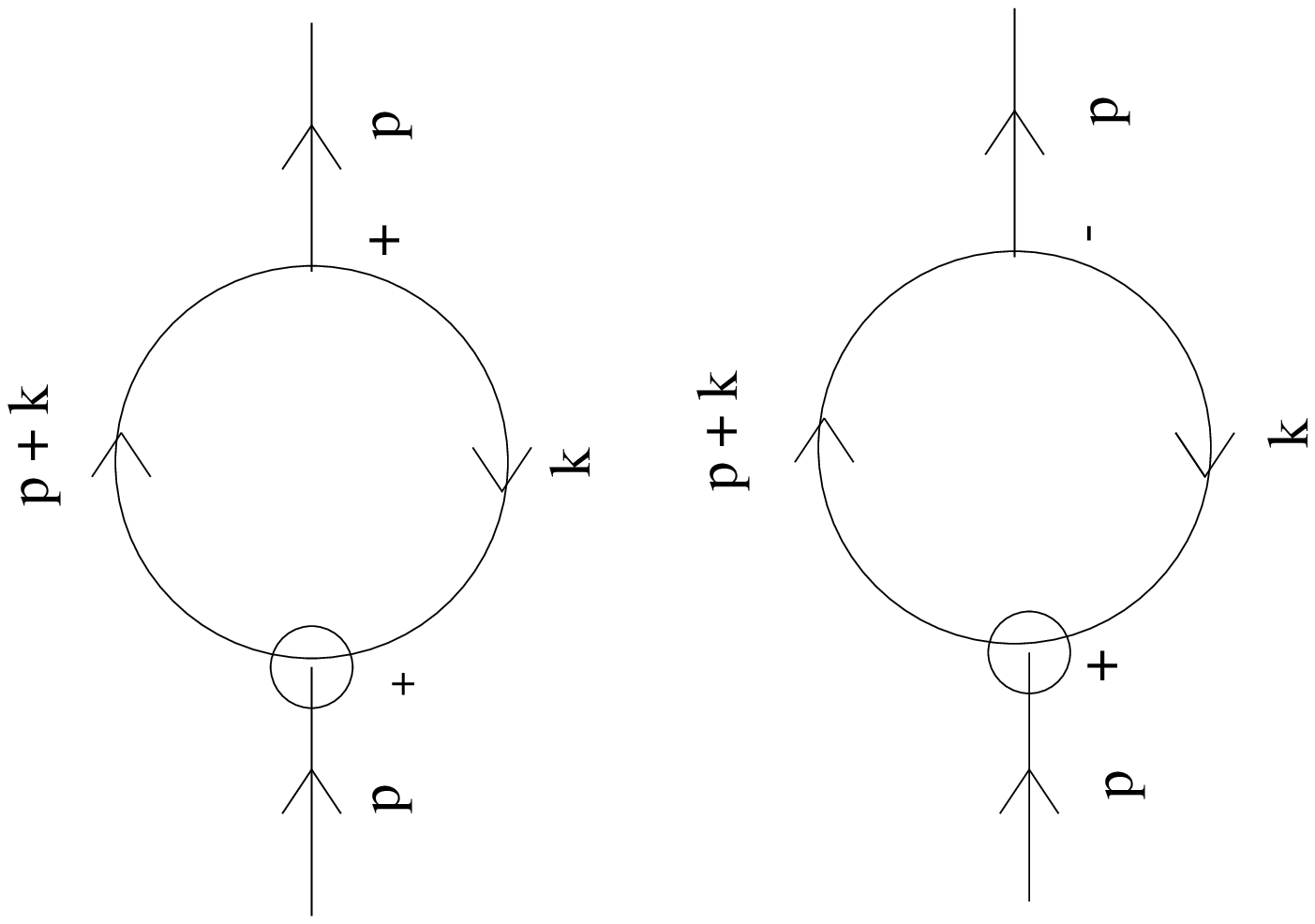,angle=270,height=7 truecm}

{\large Figure 9}
\end{center}
\bigskip\noindent
\begin{eqnarray}
\label{eq:oneloop}
i\Sigma(p)&=&-g^2 \int \dk \left[ i\D_{+-}(k)\ i\D_{-+}(p+k) 
                                -i\D_{-+}(k)\ i\D_{+-}(p+k)\right]\nonumber\\
          &=&g^2\int\dthreek {1\over 2\omega_1 2\omega_2} \big[
                 \delta(p_0+\omega_1-\omega_2)\big( n_1(n_2+1)- n_2(n_1+1)\big)\nonumber \\
          &&\phantom{g^2\int\dthreek {1\over 2\omega_1 2\omega_2} }
           +\delta(p_0-\omega_1-\omega_2)\big( (n_1+1)(n_2+1)- n_2 n_1+1\big)\nonumber\\
          &&\phantom{g^2\int\dthreek {1\over 2\omega_1 2\omega_2} }
           +\delta(p_0+\omega_1+\omega_2)\big( n_1 n_2- (n_2+1)(n_1+1)\big)\nonumber\\
          &&\phantom{g^2\int\dthreek {1\over 2\omega_1 2\omega_2} }
           +\delta(p_0-\omega_1+\omega_2)\big( n_2(n_1+1)- n_1(n_2+1)\big)\big],
\end{eqnarray}
with $\omega_1 (\omega_2)$ representing $\omega_k$ ($\omega_{p+k}$) and 
$n_1 (n_2)$ representing  $n(\omega_k) (n(\omega_{p+k}))$ respectively. This
one loop result has been obtained in several ways for various intermediate
states (as we have checked also) and
the physical interpretation of this result was already pointed out in 
\cite{ref:weldon}.
As in the zero temperature case, the  imaginary part of finite temperature 
self energies are related
to decay probabilities of the incoming particle, but with two differences. First
the probabilities of (boson) emission are enhanced by a factor ($n+1$) dependent
on the occupation number of that particle in the medium (stimulated emission)
\footnote{Pauli blocking would appear if the intermediate states included 
fermions}.
Second, new processes which do not have an analog in the vacum arise at finite 
temperature.
These are  processes involving the absorption of one or more real particles from
the medium. The absorption probability, of course, is proportional to $n$ which
represents the density of such particles present in the medium.

The possibility of extending this interpretation for higher loop graphs clearly
depends crucially
on whether we can write the imaginary part of the self-energy as
a sum of cut graphs. Only those are of the form of a decay amplitude
times its complex conjugate {\footnote{ Remember that the complex conjugate
of a graph with no circled vertices is the same graph with all vertices circled.}},
weighted by statistical factors.
That is 
exactly
what we have proved in the previous section. A simple example of how this 
works in
a two loop graph
is given pictorially in figure 10.
\bigskip
\begin{center} 
\epsfig{file=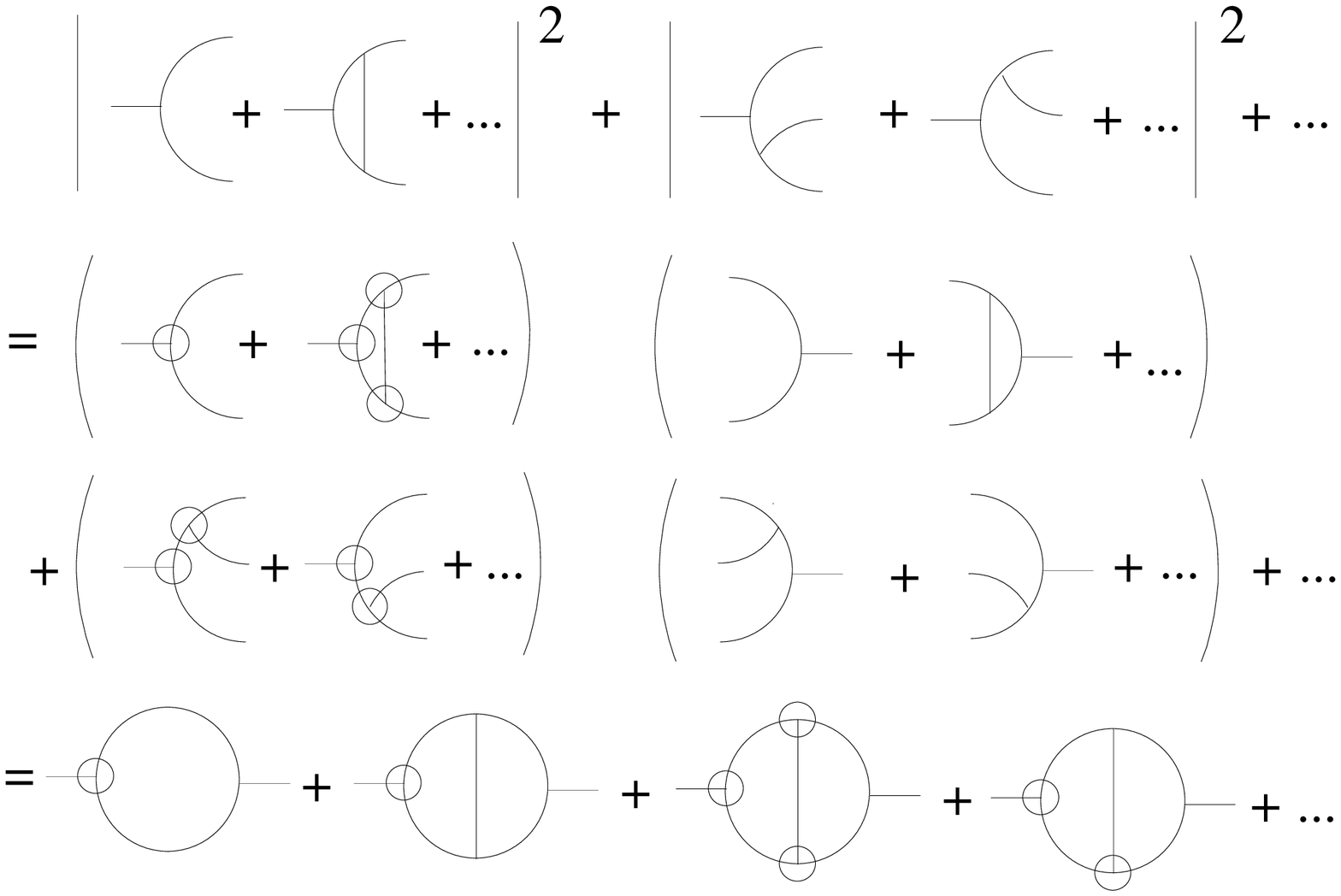,height=10 truecm}

{\large Figure 10}
\end{center}
\bigskip\noindent
 
The sum of the probabilities for the decay of the incomimg particle in 
$2$, $3$
or more particles ( plus the processes involving absorption of the 
particles from the medium ) equals the sum of the circlings in 
figure 6 that can be 
drawn as a cut diagram. The remaining graphs in figure 6 
would spoil the physical  interpretation
but they all 
vanish by our general arguments. This can also be checked explicitly. The graph in 
figure~6(l), for example, with the right, left and bottom vertices
of the type $+$ but the index of the top vertex summed over
is given by

\begin{eqnarray}
{\rm graph~ 4(l)}&=&\Delta_{++}(x-z')\Delta_{++}(y-z')\big[
                    \Delta_{+-}(x-z)\Delta_{+-}(z'-z)\Delta_{++}(y-z)\nonumber\\ 
                 &&\phantom{\Delta_{++}(x-z')\Delta_{++}(y-z')\big[ }
                   -\Delta_{+-}(x-z)\Delta_{+-}(z'-z)\Delta_{++}(y-z)\big]\\
                 &=&0,\nonumber
\end{eqnarray}
where $x,y,z,z'$ are, respectivelly, the position  of the left, right, 
top and bottom vertices. This is a special case of our first main result.
 As an example of a diagram that vanishes by our second main result, let us take
the graph in figure 4(j). Fixing, for instance,  the circled vertices to be of type $+$,
and denoting the momentum flowing from the bottom vertex to the one on the left
by $k_1$, the momentum flowing from the bottom vertex to the one on the right
by $k_2$, and the incoming momentum by $p$, the sum over the bottom vertex gives

\begin{eqnarray}
{\rm graph~ 4(j)}&=&\Delta_{--}(p+k_1)\Delta_{--}(p-k_2)\big[
                 \Delta_{+-}(k_1)\Delta_{+-}(k_2)_{+-}\Delta_{+-}(-k_1-k_2)\nonumber\\
             &&\phantom{\Delta_{--}(p+k_1)\Delta_{--}(p-k_2)\big[ }
               -\Delta_{-+}(k_1)\Delta_{-+}(k_2)_{-+}\Delta_{-+}(-k_1-k_2)\big]\\
             &=&\Delta_{--}(p+k_1)\Delta_{--}(p-k_2)
                \Delta_{+-}(k_1)\Delta_{+-}(k_2)_{+-}\Delta_{+-}(-k_1-k_2)\nonumber\\
             &&\qquad\qquad\qquad\times ~ (e^{k_1^0+k_2^0- k_1^0-k_2^0}-1)\nonumber\\
             &=&0.\nonumber
\end{eqnarray} 
\section{Conclusion}

We have considered the generalization of diagramatic cutting rules 
to the finite temperature
case using the real time formalism. Graphs that can not be represented
by cut diagrams are shown to cancel, and those that are left have a nice interpretation
in terms of decay, absorption and emission probabilities.
We have concentrated on self-energy graphs but the analysis remain
virtually unchanged for graphs with more exeternal lines.

\section{Acknowledgements}
One of us (A.D.) would like to thank Prof. H. Mani and the members of the 
Physics
group at the Mehta Research Institute, for their hospitality, where this work
started. This work was supported in part by U.S. Department of Energy grant
Nos. DE-FG-02-91ER40685 and DF-FC02-94ER40818.

\end{document}